# First results of the LARES 2 space experiment to test the general theory of relativity*

**Ignazio Ciufolini[1], Claudio Paris[2], Erricos C. Pavlis[3], John Ries[4], Richard Matzner[5], Antonio Paolozzi [2], Emiliano Ortore[2], Giuseppe Bianco[6], Magdalena Kuzmicz-Cieslak[3], Vahe Gurzadyan[7], Roger Penrose[8]**

1 Wuhan Institute of Physics and Mathematics, Innovation Academy for Precision Measurement Science and Technology (APM), Chinese Academy of Sciences, Wuhan 430071, China
2 Scuola di Ingegneria Aerospaziale, Sapienza Università di Roma, Rome, Italy
3 Goddard Earth Sciences Technology and Research II (GESTAR II), University of Maryland, Baltimore County, USA
4 Center for Space Research, University of Texas at Austin, Austin, USA
5 Center for Gravitational Physics, Weinberg Center, University of Texas at Austin, Austin Texas, USA
6 Agenzia Spaziale Italiana, CGS-Matera, Italy
7 Center for Cosmology and Astrophysics, Alikhanian National Laboratory and Yerevan State University, Yerevan, Armenia
8 Mathematical Institute, University of Oxford, Oxford, UK

*Paper dedicated to *John Archibald Wheeler*, a key figure of fundamental physics of the XX century and vigorous supporter of the LAGEOS 3/LARES 2 space experiment to test general relativity

## Abstract

The LAGEOS 3 (today LARES 2) space experiment was proposed in the eighties by the Physics Department and by the Center of Space Research (CSR) of the University of Texas (UT) at Austin and by the Italian Space Agency (ASI) to test and accurately measure frame-dragging, with the strong support of John Archibald Wheeler, director of the Center for Theoretical Physics of UT Austin. Frame-dragging is an intriguing phenomenon predicted by Einstein's theory of general relativity which has fundamental implications in high energy astrophysics and in the generation of gravitational waves by spinning black holes. LAGEOS 3 was reproposed in 2016 to the Italian Space Agency and to the European Space Agency as a technologically much improved version of LAGEOS 3 under the name LARES 2 (LAres RElativity Satellite 2) and then successfully launched in 2022 with the new launch vehicle VEGA C of ASI, ESA and AVIO. Today, after almost forty years since the original proposal, we report the first results of the LARES 2 space experiment to test general relativity. The results are in complete agreement with the predictions of Einstein's gravitational theory. Whereas previous results already confirmed the frame-dragging prediction, the conceptual relative simplicity of the LARES 2 experiment with respect to the previous tests with the LARES and LAGEOS satellites provides a significant advance in the field of tests of general relativity.

## 1. Introduction

Dragging of inertial frames, or frame-dragging as Einstein named it [1-3], is a phenomenon predicted by his theory of general relativity [4-10]. Frame-dragging is the dragging of the inertial frames and of the spacetime structure by a current of mass-energy, such as the rotation of a body, e.g., the rotating Earth. Frame-dragging is a phenomenon *fundamental* in high energy astrophysics and cosmology, not only in the theory of formation and alignment of the huge jets of plasma from active galactic nuclei and quasars by a central rotating black hole [2] but, in particular, frame-dragging has a fundamental role in the emission of gravitational waves by the coalescence of spinning black holes [11] and possibly in the misaligned accretion disk in the center of galaxy M87 [12]. Frame-dragging is a *mysterious* phenomenon, indeed around spinning black holes, with huge frame-dragging effects, and in some rotating cosmological models derived by Kurt Gödel [13,14,7], the great mathematical logician and friend of Albert Einstein, it is in principle possible to go back in time following some special time-like world-lines. Finally, frame-dragging is a *puzzling* phenomenon since some alternative theories of gravitation, e.g., the gravitational theory of Chern-Simons, which may provide an explanation to the accelerated expansion of the universe and to dark-energy and quintessence, and is equivalent to String theories of type 3, predict a value of frame-dragging different from general relativity [15] and there exists [16] also a new PPN (Parametrized Post Newtonian [17-20]) parameter measuring a value for frame-dragging different than that predicted by general relativity. Modified gravitational theories are among the approaches to deal with cosmological tensions (see, e.g. [21]). We report here a preliminary first evidence of frame-dragging based on the analysis of the LARES 2 space experiment.

With the successful launch of LARES 2 on 13 July 2022, the original idea of using two satellites in supplementary orbits to measure frame-dragging, or the so-called Lense-Thirring effect [22-28] with an accuracy well below a few percent has finally materialized. A reanalysis, after a few years of satellite laser ranging (SLR) data will be accumulated, will make possible to reach an unprecedented accuracy of its measurement that will improve by approximately one order of magnitude the best tests obtained so far [29, see also 30]. The LARES 2 satellite was injected into orbit with outstanding accuracy by the inaugural flight of the VEGA C launcher, an empowered version of VEGA which 10 years earlier placed in a lower orbit the LARES satellite. Those two launchers were developed by ASI (Italian Space Agency), ESA and AVIO, and complete the fleet of Arianspace launchers, which support the increasing demand for small payload launches. Both LARES and LARES 2 are fully supported by ASI through one of the most important laser ranging stations of the International Laser Ranging Service (ILRS), the Matera Laser Ranging Observatory (MLRO), located in Matera, Italy. It is notable that MLRO was the first laser ranging station to receive (on July 16, 2022) laser pulses reflected off LARES 2. Currently all the stations of ILRS are tracking LARES 2, providing the normal points to the scientific community (see section 3).

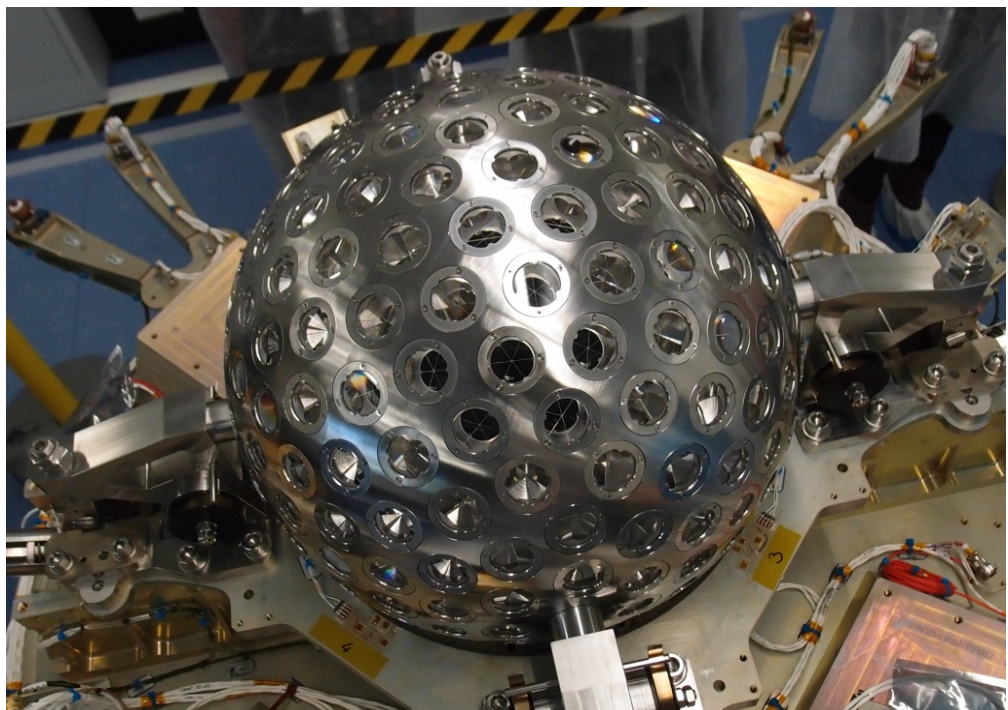

**Fig. 1. The LARES 2 satellite before launch**

## 2. A first test of frame-dragging with the LARES 2 satellite

The idea of the LARES 2 satellite, displayed in Fig. 2, was published between 1984 and 1989 in a series of papers, dissertations and NASA, ASI and CSR studies [22-28]. In the original proposals, the satellite was called LAGEOS 3 and only a few years ago was renamed LARES 2. This space experiment has the same orbit proposed for LAGEOS 3 but an improved design and structure with respect to the LAGEOS satellites [31-34]. It has indeed a much smaller cross-sectional area-to-mass, which allows to minimize the measurement biases due to the non-gravitational orbital perturbations and it is covered with smaller one-inch reflectors, allowing a sub millimeter ranging accuracy. It is then the laser-ranged satellite with the best ranging accuracy, thanks to its special distribution, and the satellite with the smallest cross-sectional area to mass, second only to LARES. (LARES has a much smaller orbital semimajor axis.) LARES 2 is a spherical satellite made of an alloy of Nickel, with a radius of 0.212 m, a mass of 294.8 kg, a cross-sectional area over mass of only 0.00047897 $m^2$/kg and is covered by 303 retroreflectors (CCRs). LARES 2 is extremely well observed by the stations of ILRS. The orbital elements of LARES 2 are: orbital inclination = 70.1615 degrees, semimajor axis = 12266.1359 km and orbital eccentricity 0.00027.

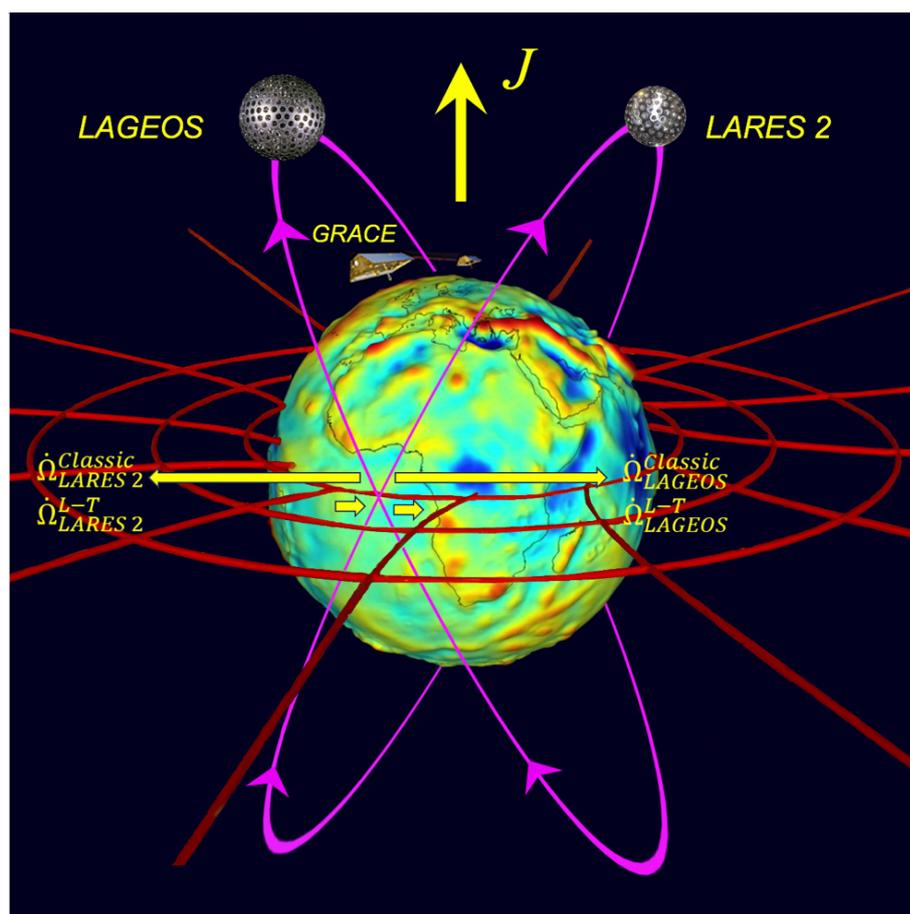

**Fig. 2. The concept of the LARES 2/LAGEOS 3 space experiment for the accurate test of frame-dragging.** The central figure of the Earth represents the Earth gravity field anomalies obtained by the space mission GRACE. By combining the orbital residuals of LARES 2 and LAGEOS is possible to eliminate [22-28] the larger shift due to the Earth's even zonal harmonics [34], including its quadrupole moment and to accurately test frame-dragging predicted by Einstein's general theory of relativity [31-34].

The concept of the LARES 2/LAGEOS 3 space experiment is displayed in figure 2. By orbiting a satellite with inclination supplementary to that of the LAGEOS satellite but with the same semimajor axis, it is possible to eliminate all the large secular perturbations on the node of the LAGEOS and LARES 2 satellites due to the Earth even zonal harmonics and therefore to be able to accurately measure frame-dragging which produces a tiny shift in their nodal longitude. The orbital inclination of a satellite is the angle between its orbital plane and the equatorial plane of the Earth, the nodal line is the intersection of the orbital plane of a satellite with Earth equatorial plane and its longitude is measured from the vernal equinox or gamma point. The even zonal harmonics describe the deviations of the Earth shape which are both symmetrical with respect to the Earth's equatorial plane and to its symmetry axis. In the spherical harmonic expansion of the Earth's gravitational potential the even zonals, $J_{2n}$, are the ones with even degree and zero order. On Earth's orbiting satellites, the even zonals produce the only secular perturbations of their nodal longitude, in addition to the much smaller frame-dragging effect. Especially large is the secular nodal precession induced by the Earth's quadrupole moment $J_2$.

Today the LARES 2 satellite may allow a test of frame-dragging with an uncertainty as small as a few parts in one thousand. This dramatic improvement with respect to the accuracy in the test of frame-dragging achievable in the eighties is due to a number of reasons. I. The GRACE and GRACE Follow-On space missions have dramatically improved our knowledge of the Earth gravitational field, both of its static and time-dependent parts. II. The new launch vehicle VEGA C of the Italian Space Agency, European Space Agency and AVIO, has injected the LARES 2 satellite into an orbit supplementary to that of LAGEOS with an exceptional precision (much better than expected on the basis of the theoretically estimated orbital injection capability of VEGA C). III. A number of new techniques, measurements and space missions have substantially improved the knowledge of the Earth tides. IV. The much smaller cross-sectional to mass ratio of LARES 2 with respect to LAGEOS and other laser-ranged satellites (with the exception of LARES) minimizes its orbital non-gravitational perturbations whereas the non-gravitational perturbations of the LAGEOS satellite have been accurately studied thanks to almost half a century of its observations. V. The LARES 2 satellite is a single spherical block of metal, thus minimizing thermal drag with respect to the LAGEOS satellites (see section 3). VI. The LARES 2 smaller one-inch retroreflectors allow better laser ranging precision than any other laser-ranged satellite. VII. The circular orbit of LARES 2 with an exceptionally small orbital eccentricity of 0.00027 implies a very small average of most of the non-gravitational perturbations of its nodal longitude due to direct radiation pressure. VIII. SLR observations of other SLR satellites, launched after the first proposal of LAGEOS 3, with extensive studies of their orbits, will also improve the modeling of the LARES 2 orbital perturbations.

A number of estimates-[22-28, see also 31-34], showed that, using the LARES 2 and LAGEOS observations, it will be possible to reach an accuracy of a few parts in one thousand in the test of frame-dragging. The price to pay to be able to reach such small accuracy in such a test and other tests of general relativity is the need to wait for a relatively long period of observations of both LAGEOS and LARES 2. In particular, for a meaningful test of frame-dragging, with improved accuracy with respect to previous tests achieved with LARES, LAGEOS and LAGEOS 2, it is necessary to wait for a period of SLR observations which is at least a multiple of the nodal period of LARES 2 and LAGEOS of about 1051 days. This is due to the need to average the effect of the Earth's lunar and solar tides [23-25,27,36,37]. The Moon and the Sun deform the shape of the Earth and the corresponding change of the Earth gravitational potential perturbs the orbits and the nodal lines of its satellites such as LARES 2 and LAGEOS.

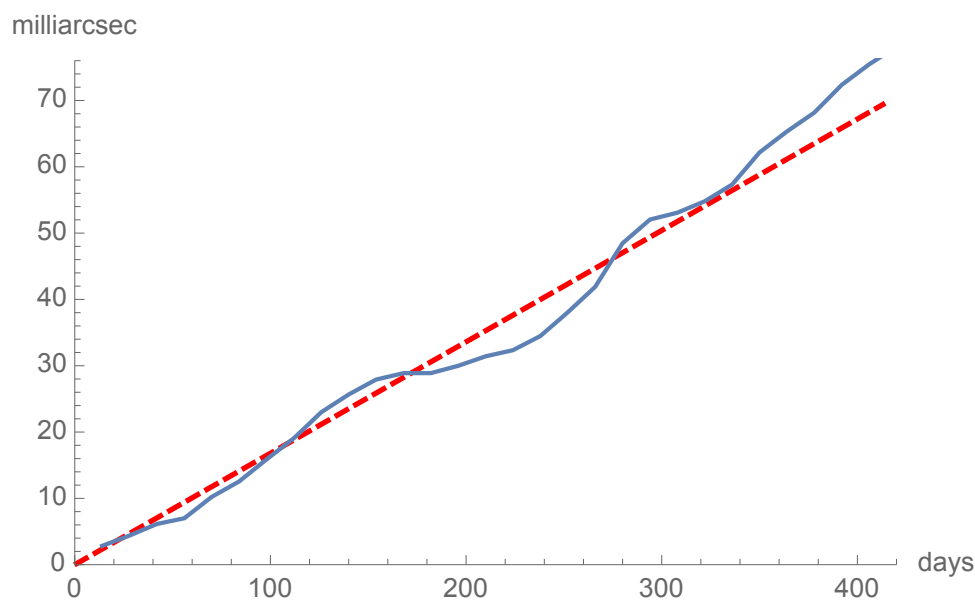

**Fig. 3 A first combination of the residuals of the nodal longitudes of LARES 2 and LAGEOS.** Obtained with 434 days of SLR data (~ 40% of the nodal period of the two orbits) after the successful launch of LARES 2, the figure shows the combined effect of frame-dragging on LARES 2 and LAGEOS (black solid line). In spite of the short period of satellite laser ranging data, the displayed drift of the nodes of LARES 2 and LAGEOS is in excellent agreement with the prediction of Einstein's general theory of relativity represented here by the red dashed line.

Nevertheless, in spite of the need to wait for a period of at least three years (precisely 1051 days), for the most statistically significant test of frame-dragging, we use a new technique [38] to eliminate part of the main effect of the Earth's tides and measure the first meaningful *combination* of the shift of the ascending nodal longitudes of LARES 2 and LAGEOS due to frame-dragging, as shown in Fig. 3.

Indeed, in this figure, almost forty years after the proposal of LAGEOS 3/LARES 2, we are able to show the first combination of the residuals of the nodal longitudes of LARES 2 and LAGEOS obtained from July 17, 2022, up to September 24, 2023, after the successful launch of LARES 2 of July 13, 2022. Let's stress the relevant features and implications of this figure. The combined shift of the nodes of LARES 2 and LAGEOS predicted by general relativity is of about 61.36 milliarcsec/yr and therefore the trend shown in Fig. 3 is in very good agreement with the general relativity prediction of frame-dragging. Such combination of the residuals of the nodal longitudes of LARES 2 and LAGEOS is unaffected by the much larger secular nodal shift due to the even zonal harmonics $J_{2n}$ and especially by the Earth's quadrupole moment $J_2$. In fact, thanks to the high level of supplementarity of the orbits of LARES 2 and LAGEOS (due to the extremely high injection precision of LARES 2 into an orbit exactly supplementary to that of LAGEOS), only an extremely small fraction of the observed effect displayed in Fig. 3 is due to $J_2$ and to the other $J_{2n}$ [35], estimated to be less than $10^{-3}$ of frame-dragging. Nevertheless, the trend displayed in this figure is affected by the uncertainties in some large Earth's tides with longer periodicity [35].

In conclusion, even though the nodal shift of frame-dragging shown in Fig. 3 is in very good agreement with the prediction of general relativity and with the previous tests of frame-dragging obtained with LARES, LAGEOS and LAGEOS 2, and in spite of the fact that the displayed trend shown in this figure was obtained by eliminating some large tidal effects, to obtain the ultimately most accurate test of frame-dragging using LARES 2 and LAGEOS, we need to wait for a much longer period of time of at least three years.

## 3. The LARES 2 satellite and other laser-ranged satellites for tests of fundamental physics, space geodesy and geodynamics

There are many differences between LARES 2 and its predecessors, LARES and the two LAGEOS satellites. LAGEOS 1 and LAGEOS 2 are constructed of three pieces held together by a long metal rod of a copper alloy. This design was chosen in the seventies (for LAGEOS) and at the end of the eighties (for LAGEOS 2) because the two external hemispherical shells of these two satellites were made of the most commonly used aerospace material: an aluminum alloy. In fact, Al alloys can be easily machined to the required tolerance and their properties in space applications are very well known. Nevertheless. the density of 2700 kg/m³, which is usually an advantage for aerospace constructions, is not acceptable in the case of laser geodetic satellites, since the effect of the non-gravitational perturbation on the orbit of a light satellite would be too large for the accuracy required to model its orbital perturbations, not only for the tiny measurements necessary for general relativity but also for geodesy, geodynamics and global climate change monitoring. Specifically, what is relevant in this class of satellites is a very low cross sectional-to-mass ratio so that they approach what, in theoretical physics, is called a test-particle. For a spherical homogeneous satellite this ratio is inversely proportional to the product of the density and the radius of the satellite. That means that a larger and denser satellite will be the most appropriate choice. Nevertheless, there is another critical parameter to consider for the design: the total mass of the satellite. In the table below all these parameters are compared.

|  | Year of launch | Alloy | Diameter (m) | Mass (kg) | Cross-Section ($m^2$) | Cross-Section/Mass ($m^2$/kg) | Number of retroreflectors |
|---|---|---|---|---|---|---|---|
| LAGEOS | 1976 | Al + Cu | 0.6 | 407 | 0.2827 | 0.00069459 | 426 |
| LAGEOS 2 | 1992 | Al + Cu | 0.6 | 405.4 | 0.2827 | 0.00069734 | 426 |
| LARES | 2012 | Tungsten | 0.364 | 386.8 | 0.1041 | 0.00026913 | 92 |
| LARES 2 | 2022 | Inconel 718 | 0.424 | 294.8 | 0.1412 | 0.00047897 | 303 |

The mass budget for a satellite is of course the primary constraint given by the launch vehicle and so, given the allowable total mass, the optimal solution for the minimization of the surface-to-mass ratio is to choose the material with the highest density. In the case of the two LAGEOS satellites, besides the two shells mentioned earlier, a copper alloy was chosen as a third component of the satellite bus: a cylinder that with its density of almost 9000 kg/$m^3$ partly compensated the low density of the aluminum alloy. However, for the two LARES satellites a more straightforward solution was chosen by abandoning completely the aluminum alloy and using a tungsten alloy for LARES and a Nickel alloy for LARES 2 [39,40]. Furthermore, these two satellites were machined out of single pieces of metal to reduce the thermal drag, a tiny but not negligible perturbation [34,41,42] as shown also by some tests performed on the LARES satellite. In contrast to the other three satellites, LARES 2 has adopted Commercial Off The Shelf (COTS) retroreflectors with a smaller front face diameter: 1 inch instead of 1.5 inches. Another peculiar characteristic of LARES 2 is the positioning of the Cube Corner Reflectors (CCRs): no longer regularly distributed along each single parallel of the satellite [38], the CCR locations are obtained from the best-known solution to the Thomson's Problem for 303 CCRs, properly re-adapted to accommodate the interfaces with the separation system of the satellite. (The separation interface maintains the satellite on the launcher and is designed to withstand the accelerations and the vibrations during the launch and to release the satellite once the final orbit is reached. The separation system and relevant interfaces are very similar to those of LARES.) The unique CCR distribution used on LARES 2 reduces one of the ranging measurement errors, the *satellite signature*, to below 1 mm. The choice of commercial rather than custom made CCRs is a novelty for geodetic laser-ranged satellites. To mitigate the risk involved with this choice a series of tests, used also to qualify LARES CCRs, was repeated on a sample of ten LARES 2 COTS CCRs. These laboratory measurements are now confirmed by the quality of the actual laser ranging data acquired for LARES 2 during the 15 months (so far) of LARES 2 in orbit. Figure 4 shows the size of the ranging residuals of the Matera station for LARES 2.

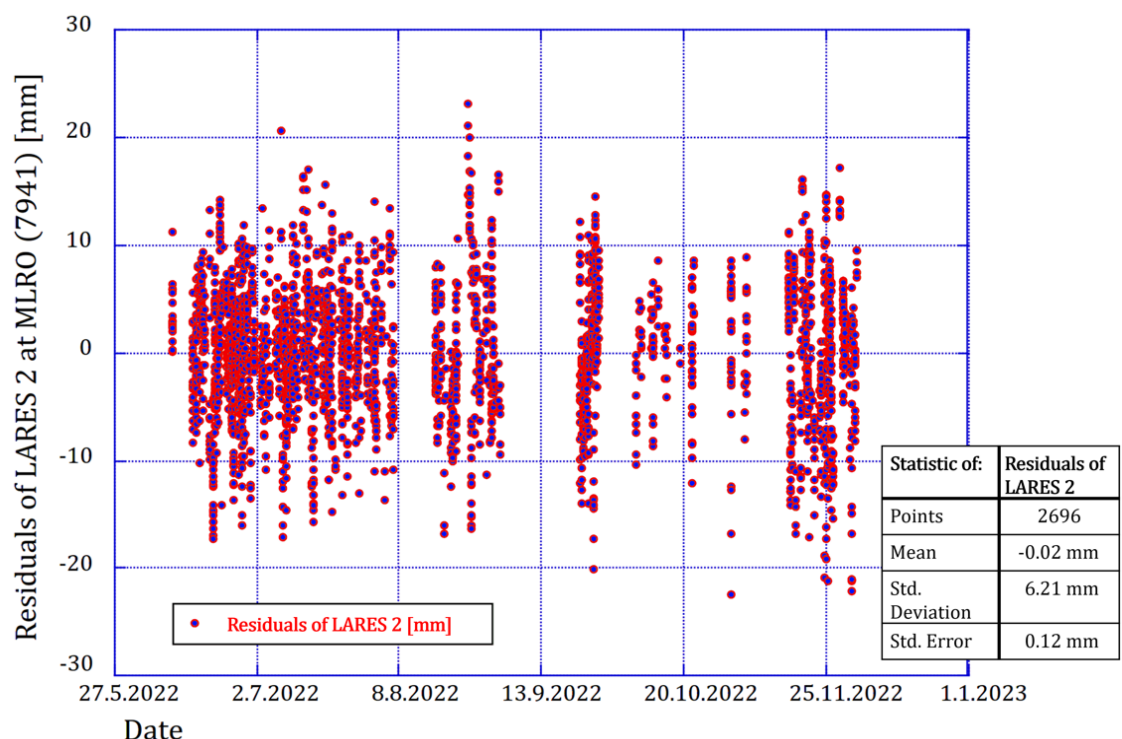

**Figure 4. LARES 2 range residuals of the Matera Laser Ranging Observatory**. The y-axis (range residuals) units are millimeters.

## 4. Conclusions

The first results of the LARES 2 space experiment are reported here. In spite of the fact that a much longer period of satellite laser ranging data is needed (to eliminate the biasing effect of the Earth's non-zonal harmonics due to long period tides) for an ultimately accurate test and measurement of frame-dragging, we have been able to derive a first preliminary combination of the LARES 2 and LAGEOS data. Our results fully confirm the prediction of frame-dragging in Einstein's general theory of relativity, a phenomenon which has a basic role in high energy astrophysics and in the formation of gravitational waves by spinning black holes.

**Acknowledgements** The authors wish to thank the Italian Space Agency for having funded the design of LARES 2 satellite under agreement n. 2017-23-H.0, the European Space Agency and AVIO for the VEGA C inaugural flight and the International Laser Ranging Service [43] for tracking the satellites and providing the laser ranging data. E.C. Pavlis acknowledges the support of NASA Grant 80NSSC22M0001 and computational resources provided by NASA's High-End Computing (HEC).

**Data Availability Statement** Satellites Laser Ranging (SLR) data of LARES 2 and LAGEOS are available at the NASA CDDIS (Crustal Dynamics Data Information System) as well as the ILRS European Data Center (EDC).